\documentclass[intlimits,twoside,a4paper]{article}

\usepackage[cp1251]{inputenc}
\usepackage[eqsecnum]{cmpj3}

\usepackage{bm}

\issue{2023}{26}{2}{23603}
\doinumber{10.5488/CMP.26.23603}

\title[Development of a new force field]
      {Development of a new force field for the family of primary aliphatic
        amines using the three steps systematic parameterization procedure}

      \author[H. Espinosa-Jim\'enez, A. B. Salazar-Arriaga,
        H. Dominguez]
{H. Espinosa-Jim\'enez, A. B. Salazar-Arriaga,
  H. Dominguez\orcid{0000-0001-6126-9300}\footnote{Corresponding author: \email{hectordc@unam.mx} }}

\address{Instituto de Investigaciones en Materiales, 
Universidad Nacional Aut\'onoma de M\'exico, UNAM Cd. Mx. 04510, M\'exico}


\date{Received July 22, 2022, in final form January 18, 2023}

\begin{document}

\maketitle

\begin{abstract}

  The applicability of the three steps systematic parametrization procedure
  (3SSPP) to develop a force field for primary amines was evaluated in the
  present work.
  Previous simulations of primary amines show that current
  force fields (FF) can underestimate some experimental
  values under room conditions. Therefore, we propose a
  new set of parameters, for an united atom (UA) model,
  that can be used for short and long amines which predict correctly
  thermodynamic and dynamical properties.
  Following the 3SSPP methodology, the partial charges
  are chosen to match the experimental dielectric constant
  whereas the Lennard-Jones (LJ) parameters, $\epsilon$ and $\sigma$, are
  fitted to reproduce the surface
  tension at the vapor-liquid interface and the liquid density, respectively.
  Simu\-lations were initially
  conducted for the propylamine molecule by
  introducing three different types of carbon atoms, C$_\alpha$ and C$_\beta$,
  with electric charges, and C$_n$, without charge. Then, modifying the
  charges of the carbons and using the transferable LJ parameters, the new
  set of constants for long amines were found. The results
show good agreement for the experimental dielectric constant and mass density 
with a percentage error less than 1\%, whereas for the
surface tension the error is up to 4\%. For the short amines, methylamine and
ethylamine, the new charges were obtained from a fitting function calculated
from the long amines results. For these molecules, the values of
the dielectric constant and the surface tension present errors
of the order of 10\% 
with the experimental data. Miscibility of the amines was also tested
with the new parameters and the results show reasonable agreement
with experiments.
\printkeywords
%
%
\end{abstract}


\section{Introduction}

Amines are molecules derived from ammonia with one or more alkyl
or aryl groups with great interest in areas of chemical engineering.
They are used in several industrial
applications, such as carbon dioxide retainers~\cite{1} and to remove hydrogen
sulfides and carbon dioxide from natural gas~\cite{2,3}.
Due to their biological activity they have also been used in drugs and
medicines~\cite{4,5}.
All of these attributes make amines very interesting molecules to study
not only from the experimental but also from the theoretical
and computational points of view.

Nowadays, computer simulations have become an important tool to
study complex systems. 
In particular, molecular dynamics simulations are a good alternative
to understanding and  obtaining complementary information that can be difficult
to collect from a laboratory. However, in order to have 
reliable computational results it is necessary to have good force
fields that can reproduce
several thermodynamic, dynamic and structural properties.

For the case of amines, several properties were
predicted using different force fields. For instance,  
Rizzo et al. used an all atom (AA) OPLS model to report the liquid density
and enthalpy of vaporization at one temperature showing
reasonable agreement with experiments~\cite{6}.
Wick et al. used a different AA-model, TraPPe-EH, to conduct simulations
of amines at different temperatures above the boiling point~\cite{7}.

A different approach to the AA model is to consider
the united atom (UA) model where each chemical group is reduced to
one single site with
appropriate parameters. In particular, a few years ago the UA model
with shifted Lennard-Jones centers (AUA)~\cite{8}, and
reparametrized time after (AUA4)~\cite{9}, was used to study linear and
branched hydrocarbons with good results to predict fluid densities
and pressures~\cite{10,11,12,13}.
Those models were also tested for alkylamides
and alkanols giving partial agreement with actual experiments~\cite{14}.
Brad and Patel, studied several thermodynamic and dynamic properties
of methylamine and ethylamine using charge equilibration force fields~\cite{15}
and they report discrepancies above 15\% in some properties, e.g.,
enthalpy of vaporization.

On the other hand, Orozco et al.~\cite{16} proposed an anisotropic
united atom
force field (AUA4) with four charges to study phase equilibrium of several
primary amines. They used displaced force centers for the Lennard-Jones
potential and predicted thermodynamic and transport properties with
acceptable results compared to experimental data. Later, the same authors
using the same model, studied equilibrium and transport properties of
primary, secondary and tertiary amines and they obtained good
agreement with real experiments~\cite{17}.

 In the present study, we propose a simple united atom
force field for primary
amines using the 3SSPP
method reported a few years ago~\cite{18}. The methodology was tested in
several systems with very good
results~\cite{19,20,21}, and then we used the same methodology for the primary
amine molecules.

\section{Computational model}
\label{sec2}

\begin{figure}[h]
\begin{center}
\includegraphics[scale=0.28]{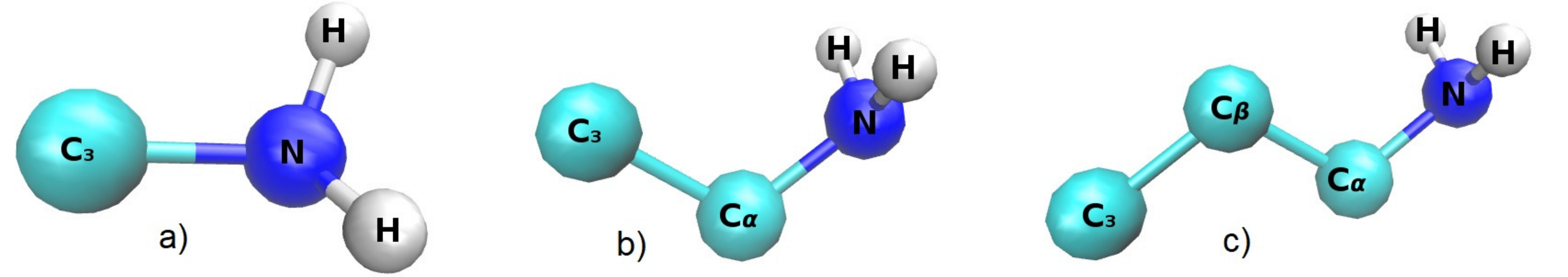}
\caption{(Colour online) Molecule structure of a) methylamine, b) ethylamine and
  c) propylamine. The atom sites (N and~H) for the NH$_2$ group and carbons
  in the tails (C$_\alpha$, C$_\beta$ and C$_n$)
  are indicated in the figures.}
\label{fig1}
\end{center}
\end{figure}

 Primary amines were constructed using an amino
group, NH$_2$, with the nitrogen and hydrogen atoms explicitly
modelled, attached to a hydrocarbon tail of united atoms for the CH$_2$ and
CH$_3$ alkyl groups. In the united
atom model, the first CH$_2$ group attached to the NH$_2$ group is named
C$_\alpha$, in the case of long amines the next CH$_2$ is named C$_\beta$,
the rest CH$_2$ are C$_2$ and the last CH$_3$ group is called C$_3$,
see figure~\ref{fig1}.

In the present work, the amine force field considers inter-molecular and
intra-molecular interactions,
\begin{equation}
U = U_{\rm bond} + U_{\rm ang} + U_{\rm dih} + U_{\rm LJ} + U_{\rm coul}.
\end{equation}
\noindent The first three elements, $U_{\rm bond}$, $U_{\rm ang}$,  $U_{\rm dih}$,
correspond to the
intra-molecular interactions (bond, angle and torsional potentials,
respectively) and the last two terms, $U_{\rm LJ}$ and $U_{\rm coul}$, represent the
inter-molecular interactions (Lennard-Jones and coulombic potentials,
respectively). Several years ago, Rizzo and Jorgensen showed good
hydrogen-bond strengths and hydration results for amines simulations
using the  OPLS force field~\cite{6}. Therefore,
for the intra-molecular interactions we chose the OPLS model as initial
force field. For the inter-molecular interactions we decided
to use the TraPPe model used by Wick et al.~\cite{7} since they
obtained good densities and critical
temperatures, i.e., bulk properties, in their simulations
of amines. On the other hand, the initial charges were
obtained with the RESP (restricted electrostatics potential) method~\cite{22}.

All simulations were conducted using the molecular dynamics
method in the isothermal-isobaric (NPT)
and canonical (NVT) ensembles. The dielectric constants, densities and
diffusion coefficients
were calculated in the NPT simulations with
500~amine molecules at constant temperature, $T = 298.15$~K, and pressure,
$P = 1$~bar. The internal temperature was coupled to a
Nos\'e-Hoover thermostat~\cite{23} with a relaxation time of $\tau_{T} = 1$~ps,
while the pressure
was coupled to a Parrinello-Rahman barostat~\cite{24} with a time parameter
of  $\tau_{P} = 1.0$~ps. The short range interactions were cutoff at 20~\AA
{} and simulations were performed up to 50~ns after 10~ns of equilibration.

The NVT simulations were used to calculate surface tensions
in a box of dimensions $L_x = L_y = 5.0$~nm and $L_z = 15.0$~nm with 1000
amine molecules in a liquid phase in the middle of the box, i.e., two
liquid-vapor interfaces
were constructed. Simulations were conducted at constant temperature,
$T = 298.15$~K, using the Nos\'e-Hoover thermostat~\cite{23} with a relaxation
time constant of  $\tau_{T} = 1$~ps. In this case,
the short range interactions were cutoff
at 25~\AA {} to avoid any size effects on this property, as recommended in
previous works~\cite{25}. Then, simulations run for 30~ns after 10~ns of
equilibration.

The simulations to measure miscibility of amines in water 
were carried out in the NPT ensemble with 4000 water molecules and 100
amines, placed initially in an homogeneous mixture, in a box of initial
dimensions of $5 \times 5 \times 5$~nm. The pressure was $P = 1$~bar and temperature $T = 298$~K,
with relaxation times of  $\tau_{P} = 1$~ps and  $\tau_{T} = 1$~ps,
respectively. The short range
interactions were cutoff at 20~\AA {} and simulations were performed up to
50~ns after 10~ns of equilibration. For these simulations, we chose a water
model which correctly reproduced several experimental data, including our
target properties, the dielectric constant and the surface tension,
TIP4P/$\epsilon$~\cite{26}.

All simulations were carried out with GROMACS-2021~\cite{27}
software applying
periodic boundary conditions in all directions using the leap-frog
algorithm~\cite{28} with a time step of $dt = 2$~fs to solve the equations of
motion with compressibility value of $4.5 \times 10^{-5}$~bar$^{-1}$. The unlike
interactions between distinct atoms were calculated
using the Lorentz-Berthelot combination rules~\cite{28} and  the electrostatic
interactions were handled with the particle mesh Ewald
method~\cite{29} (fourth order with a Fourier spa\-cing~0.16) whereas
bond lengths were constrained using the Lincs algorithm~\cite{30}.

\subsection{Optimization procedure}

The new force field was constructed following the three steps
systematic parameterization procedure developed a few years ago~\cite{18}.
In that method three experimental properties are chosen as target quantities
to be fitted, the dielectric constant, the surface tension and the liquid
density by scaling the charges and Lennard-Jones parameters of all
atoms in the molecules.

In the first step of the 3SSPP scheme all the partial charges
are scaled from the original ones to match the experimental dielectric
constant. Then, using the NPT ensemble, simulations were
conducted with a set of charges, and the static dielectric constant, $\epsilon$,
was estimated by the time average of the fluctuations of the
dipolar moment, $\textbf{M}$, of the whole system,
\begin{equation}
  \epsilon = 1 +  \frac{4\piup}{3k_{\rm B}TV}\left(\langle \textbf{M}^2 \rangle -
  \langle \textbf{M} \rangle ^2 \right),
  \label{2.2}
\end{equation}

\noindent where $k_{\rm B}$ is Boltzmann's constant, $T$ is the absolute temperature,
$V$ is the volume of the simulation cell, and $\textbf{M}$ is the
summation of the dipole moment vectors of all the atoms,
\begin{equation}
  \textbf{M} = \sum_{i}q_{i} \textbf{r$_i$},
  \label{2.3}
\end{equation}
\noindent where $q_i$ and \textbf{r$_i$} are the charge and position of
atom $i$.
The angled brackets in equation~\eqref{2.2} indicate a time average.

For those calculations,
partial charges were imposed in the nitrogen and hydrogens (of the
NH$_2$ group), C$_\alpha$ and
C$_\beta$ whereas the C$_n$ sites had zero charge.

The second step of the 3SSPP consists in parameterization of
the $\epsilon_{\rm LJ}$ Lennard-Jones parameters to fit the experimental
surface tension ($\gamma$) using the mechanical expression,
\begin{equation}
  \gamma = \frac{L_z}{2} \left[\langle P_{zz} \rangle - \frac{1}{2}
\left(\langle P_{xx} \rangle + \langle P_{yy} \rangle \right)\right].
\label{2.4}
\end{equation}
The $\langle P_{ii} \rangle$ are the components of the
stress pressure and $L_z$ is the length
of the simulation box. The factor $1/2$ is for the two liquid-vapor
interfaces in the simulation box.

In the third step, all the $\sigma_{\rm LJ}$ Lennard-Jones
parameters are scaled to obtain the experimental mass density.

\section{Results and discussion}

\subsection{Dielectric constant}

Simulations started for the propylamine molecule. Firstly,
the original partial charges for the nitrogen, hydrogen (of the
NH$_2$ group) and C$_\alpha$ sites, (the C$_\beta$ and C$_3$ charges were zero) 
were scaled until the experimental dielectric constant was fitted with an
error less than 5\%. In table~\ref{table1} the new and original charges are shown.

\begin{table}[h]
\begin{center}
	\caption{\small{Propylamine charges and Lennard-Jones parameters.
			The ``org'' means
			the original values taken from TraPPe force field~\cite{7}.}}
		\label{table1}
\begin{tabular}{|c|c|c|c|c|c|}
  \hline
  Site  & $q$ (e) & $q_{\rm org}$ ($e$) & $\sigma_{\rm LJ}$ (nm) & $\sigma_{\rm LJ, org}$ (nm) &
  $\epsilon_{\rm LJ}$(KJ/mol)\\
\hline
C$_3$     & 0.000  & 0.000 & 0.3817 & 0.375 & 0.8148\\
C$_\beta$  & 0.000  & 0.000 & 0.4021 & 0.395 & 0.3824\\
C$_\alpha$ & 0.237  & 0.180 & 0.4021 & 0.395 & 0.3824\\
N         & $-0.977$ &$-0.892$ & 0.3400 & 0.334 & 0.9229\\
H         & 0.370  & 0.356 & 0.0000 & 0.000 & 0.0000\\
\hline
\end{tabular}
\end{center}
\end{table}

As the hydrocarbon chain increases (for long amines), the NH$_2$
group modifies its polar activity, i.e., a negative charge arises in the
C$_\beta$ carbon and consequently the charge of
the C$_\alpha$ changes to keep the neutrality of the system.
Then, the C$_\alpha$ and C$_\beta$ charges of the long amines should be modified
to fit the experimental dielectric constants.

\begin{figure}[h]
	\begin{center}
		\includegraphics[width=2.5in, angle=-90]{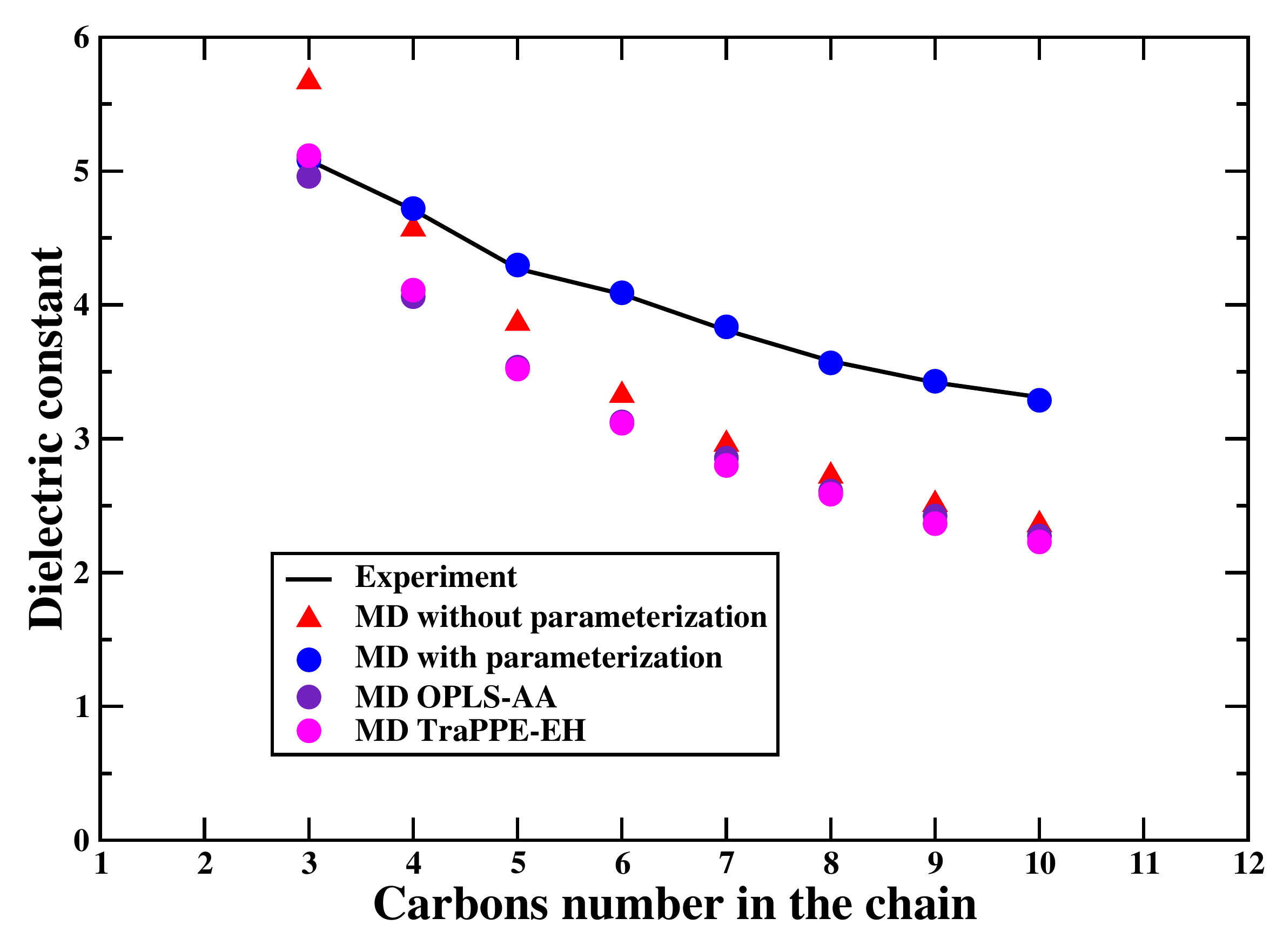}
		\caption{(Colour online) Dielectric constants of the $n$-amines as function
			of the number of carbon atoms in the alkyl chains. Data 
			calculated from this work (new and old parameterization) and compared with
			OPLS-AA~\cite{6}, TraPPe-EH~\cite{7} models
			and experiments~\cite{31}.}
		\label{fig2}
	\end{center}
\end{figure}

In table~\ref{table2}, the values of C$_\alpha$ and C$_\beta$ charges are shown
for the $n$-amines
from three (propylamine) to ten (decylamine) carbons in the alkyl
chains. The results for the dielectric constants are given in table~\ref{table3} and
figure~\ref{fig2}. In the same figure~\ref{fig2} comparisons
with other force field are included where it is observed
that the values with the new parameters have an error less
than 1\% with the experiments.

\begin{table}[h]
	\begin{center}
		\caption{\small{Charges of the different carbons (C$_3$, C$_2$,	C$_\beta$ and C$_\alpha$) in the amine molecules for all
		the amines from 1 to 10  carbons
		in the hydrocarbon tail. Charges of the system with 2 carbons were
		obtained from equation~\eqref{3.2}.}}
		\label{table2}
		\begin{tabular}{|c|c|c|c|c|}
			\hline
			Amines, num. of carbons & C$_3$ & C$_2$ & C$_\beta$ & C$_\alpha$\\
			\hline
			1  & 0.237 &   $-$    &    $-$   & $-$    \\
			2  & 0.042 &   $-$    &    $-$   & 0.195\\
			3  & 0.000 &   $-$    &  0.000 & 0.237\\
			4  & 0.000 & 0.0000 & $-0.096$ & 0.333\\
			5  & 0.000 & 0.0000 & $-0.111$ & 0.348\\
			6  & 0.000 & 0.0000 & $-0.130$ & 0.367\\
			7  & 0.000 & 0.0000 & $-0.141$ & 0.378\\
			8  & 0.000 & 0.0000 & $-0.148$ & 0.385\\
			9  & 0.000 & 0.0000 & $-0.157$ & 0.394\\
			10 & 0.000 & 0.0000 & $-0.164$ & 0.401\\
			\hline
		\end{tabular}
	\end{center}
\end{table}

\begin{figure}[h]
\begin{center}
\includegraphics[width=2.5in, angle=-90]{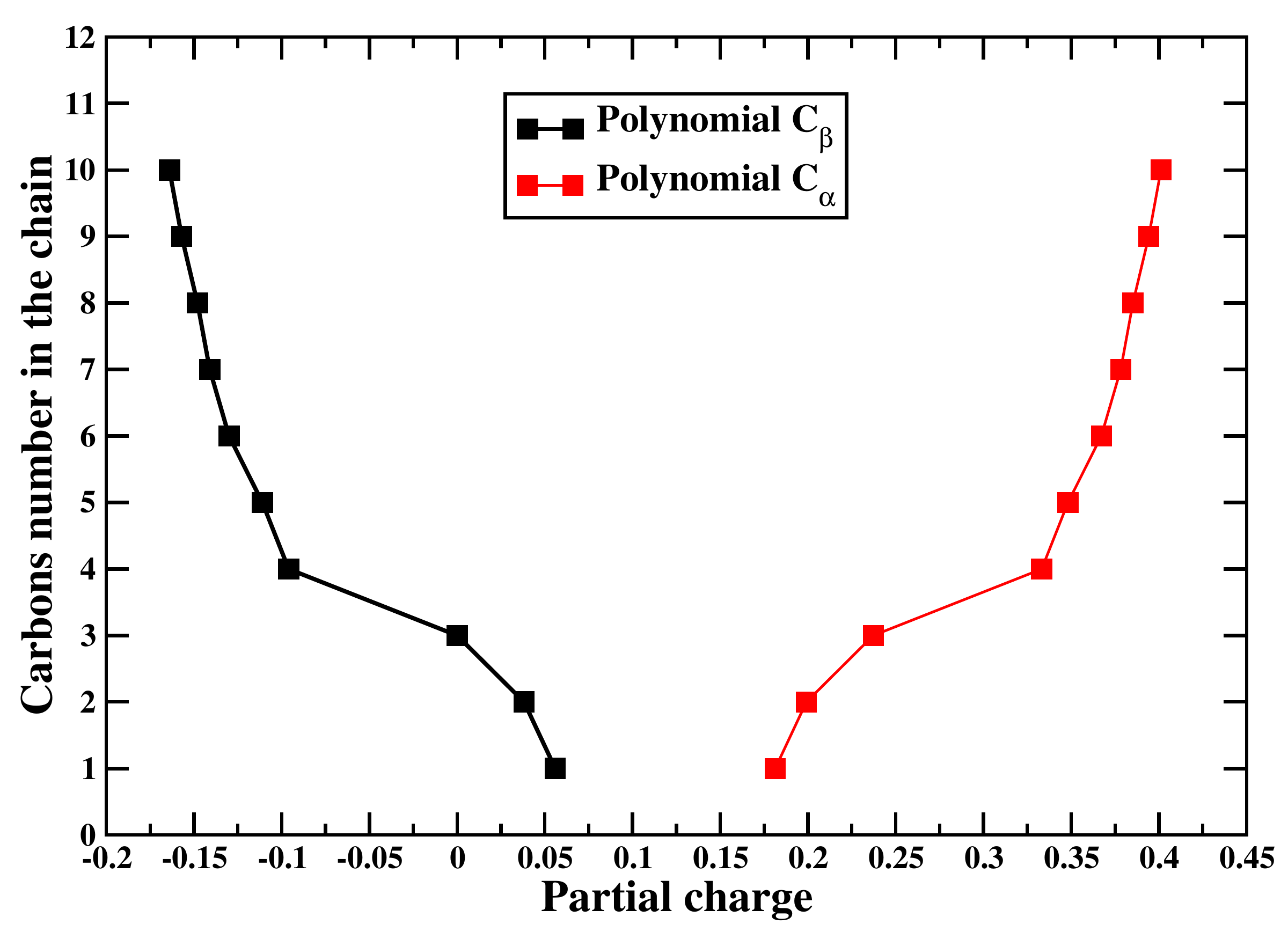}
\caption{(Colour online) Best fitting curve for the relationship between the amine
  chain length and the partial charges of C$_\alpha$
  and C$_\beta$ calculated with equations~\eqref{3.1} and ~\eqref{3.2}. }
\label{fig3}
\end{center}
\end{figure}

In figure~\ref{fig3}, a plot of the chain length (number of carbons) as
function of the C$_\beta$ and C$_\alpha$ carbon charges are shown along with the
best fitting curves
to the data. Several functions were tested to fit the data and the best
one was a third order polynomial.

For the C$_\beta$, the function that best fits the data is,
\begin{equation}
N_{\beta}(q) = -2809.1q^3 - 271.92q^2 - 11.721q + 2.9986,
\label{3.1}
\end{equation}

\noindent whereas for the C$_\alpha$ the fitting curves is,

\begin{equation}
N_{\alpha}(q) = 2809.1q^3 - 2271q^2 + 614.92q - 52.577,
\label{3.2}
\end{equation}

\noindent where $N_{\alpha,\beta}(q)$ represent the number of carbons
in the molecule.

\subsection{Parameterization of Lennard-Jones parameters}

Once the dielectric constant of the propylamine is obtained,
the next step in the  SSPP3 method is the evaluation of the surface tension.
For that calculation, the charges are fixed
and the simulations are carried out by scaling the $\epsilon_{\rm LJ}$
values of all the atoms in the molecule. However,
the surface tension did
not change significantly with any variation of the $\epsilon_{\rm LJ}$, i.e.,
the best results were obtained with the original parameter.

With the $\epsilon_{\rm LJ}$ parameters and charges, the $\sigma_{\rm LJ}$
were now modified, i.e.,
$\sigma_{\rm LJ}$ of all atoms in the molecule were scaled,
in increments of 10\%, and mass density was
calculated until the experimental data were obtained, within an error less
than 5\%. The results for all the amines are given in table~\ref{table3}.

\begin{table}[h]
\begin{center}
	\caption{\small{Data of the dielectric constant ($\epsilon$), surface
			tension ($\gamma$) and mass density ($\rho$) for all the amines.
			The table shows values calculated from this work, the actual
			experiments (subindex `exp') and the percentage error. Data
			at $T = 298$~K except for `$^*$' which was obtained at $T = 215$~K.}}
	\label{table3}
\begin{tabular}{|c|c|c|c|c|c|c|c|c|c|}
  \hline
  Amines & $\epsilon$ & $\epsilon _{\rm exp}$ & error & $\gamma$ & $\gamma _{\rm exp}$ &
  error & $\rho$ & $\rho _{\rm exp}$ & error \\
    &  &  & (\%) & (mN/m) & (mN/m) & (\%) & (kg/m$^3)$ & (kg/m$^3)$ & (\%) \\
\hline
Methylamine & 17.54$^*$& 16.7& 5.00 & 21.93& 19.88 & 10.27& 664.13 & 655 & 1.39\\
Ethylamine  & 8.62$^*$ & 8.7 & 0.86 & 21.96& 19.89 & 10.42& 720.28 & 677 & 6.39\\
Propylamine & 5.083   & 5.08 & 0.06 & 23.8 & 22.85 & 3.99 & 716.04 & 714 & 0.28\\
Butylamine  & 4.719   & 4.71 & 0.19 & 25.3 & 24.15 & 4.54 & 739.39 & 741 & 0.22\\
Pentylamine & 4.298   & 4.27 & 0.65 & 26.4 & 25.45 & 3.59 & 754.73 & 751 & 0.49\\
Hexylamine  & 4.090   & 4.08 & 0.24 & 27.3 & 26.83 & 1.72 & 767.09 & 761 & 0.80\\
Heptylamine & 3.835   & 3.81 & 0.66 & 27.9 & 27.40 & 1.50 & 776.71 & 772 & 0.61\\
Octylamine  & 3.566   & 3.58 & 0.39 & 28.5 & 28.24 & 0.91 & 784.97 & 779 & 0.77\\
Nonylamine  & 3.429   & 3.42 & 0.26 & 28.9 & 28.68 & 0.76 & 792.32 & 785 & 0.93\\
Decylamine  & 3.287   & 3.31 & 0.69 & 29.3 & 29.50 & 0.68 & 797.34 & 791 & 0.80\\
\hline
\end{tabular}
\end{center}
\end{table}

\begin{figure}[h]
	\begin{center}
		\includegraphics[width=2.7in, angle=-90]{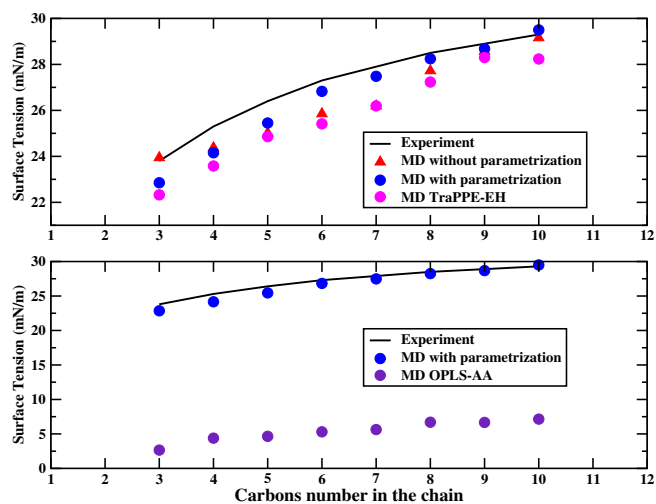}
		\caption{(Colour online) Surface tensions of the $n$-amines as function
			of the number of carbon atoms in the alkyl chains at temperature $T = 298$~K.
			Data calculated from this work (new and old parameterization)
			and compared with OPLS-AA~\cite{6}, TraPPe-EH~\cite{7} models
			and experiments~\cite{32}~(ID:~6089, 6101, 7564, 7716, 7769, 7811, 7835, 7851, 15387, 8576).}
		\label{fig4}
	\end{center}
\end{figure}

Transferability of the new $\epsilon_{\rm LJ}$ and $\sigma_{\rm LJ}$
was evaluated by simulating the rest of the long amines and considering the
charges described in the previous section.
Then, the surface tensions and densities
were calculated using the $\sigma_{\rm LJ}$ and $\epsilon_{\rm LJ}$ parameters
of the carbons in the long
amines. The surface tension results are plotted in figure~\ref{fig4}
where it is possible to see that the new reparameterization gives slightly
better agreement with the experiments~\cite{32}~(ID:~6089, 6101, 7564, 7716, 7769, 7811, 7835, 7851, 15387, 8576), although in some cases
the error is about 5\% as in the butylamine (see table~\ref{table3}).

\begin{figure}[h]
\begin{center}
\includegraphics[width=2.5in, angle=-90]{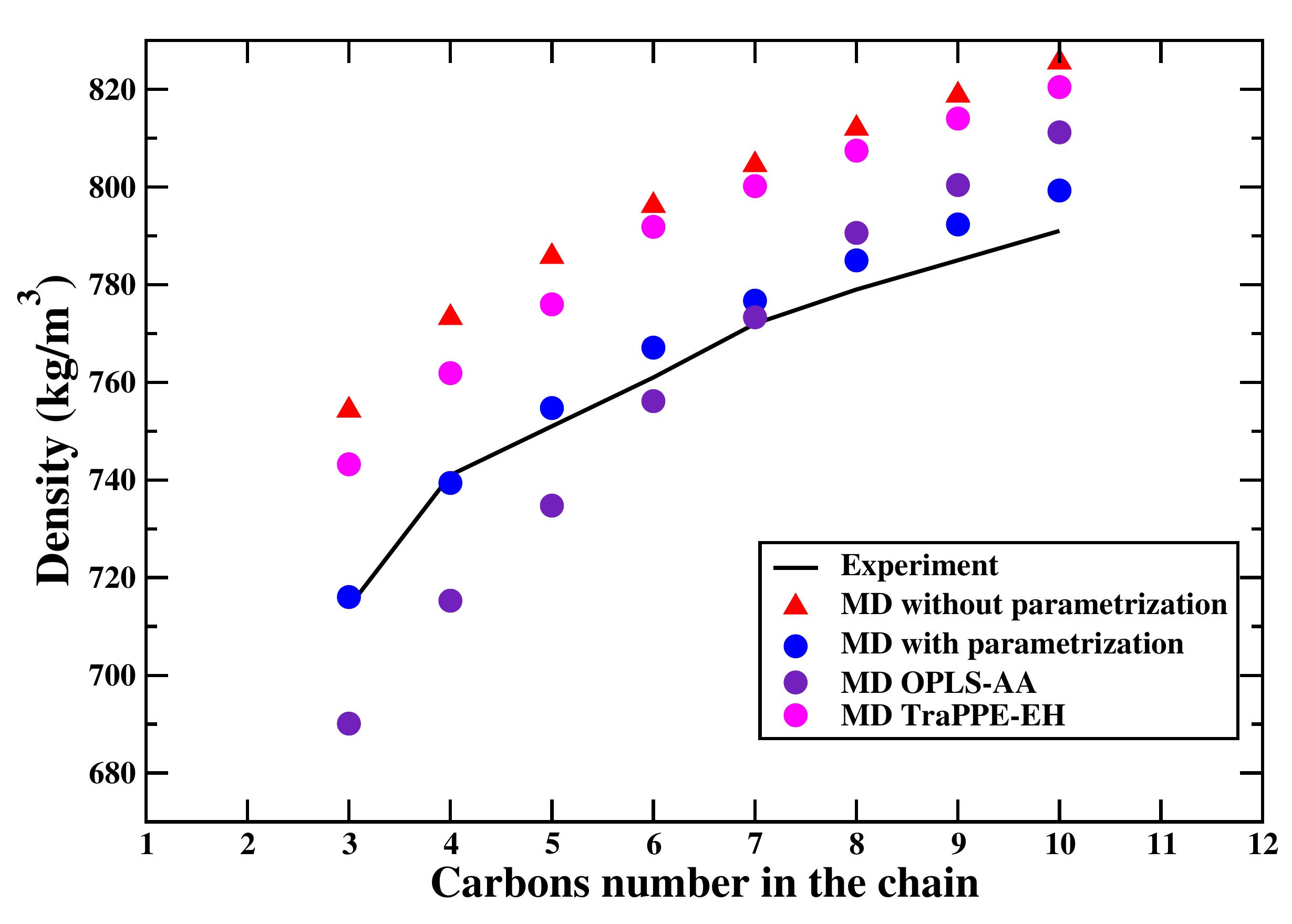}
\caption{(Colour online) Mass density of the $n$-amines as function
  of the number of carbon atoms in the alkyl chains at temperature $T = 298$~K.
  Data calculated from this work (new and old parameterization)
  and compared with OPLS-AA~\cite{6}, TraPPe-EH~\cite{7} models
and experiments~\cite{33}.}
\label{fig5}
\end{center}
\end{figure}

In figure~\ref{fig5}, the mass density data are plotted
and compared with actual 
experiments~\cite{33}. It can be observed that the new values
are in much better agreement than the results with the old parameters
and other force fields. In our case, the OPLS-AA data for
the propylamine and butylamine differ by 4\% from those reported in
references~\cite{6,33a}.

With the above results, transferability of the force field was
also tested for the shortest amines, the methylamine and ethylamine.
The methylamine molecule has only one C$_3$ site and its charge was
chosen to have a neutral charged molecule. In the case of the ethylamine,
it has carbon C$_\alpha$, where its charge was estimated from
equation~\eqref{3.2}, and carbon C$_3$, with an electric charge determined to
have a molecule with zero total charge. In table~\ref{table2}, the values of the
charges for the methylamine and
ethylamine are shown. Using those charges and the $\epsilon_{\rm LJ}$ and
$\sigma_{\rm LJ}$ Lennard-Jones parameters,
the target properties were calculated, i.e., the dielectric constant, the
surface tension and the density. In the case of the dielectric constant
we found errors of 5\% and 0.86\% with the experiments for the
methylamine and ethylamine, respectively. For the densities, the
errors were $\approx$ 1\% and 6\%, for
the methylamine and ethylamine, respectively. For the surface tension,
the errors were about 10\% in both molecules, see table~\ref{table3}.

\begin{figure}[h]
	\begin{center}
		\includegraphics[width=2.7in, angle=-90]{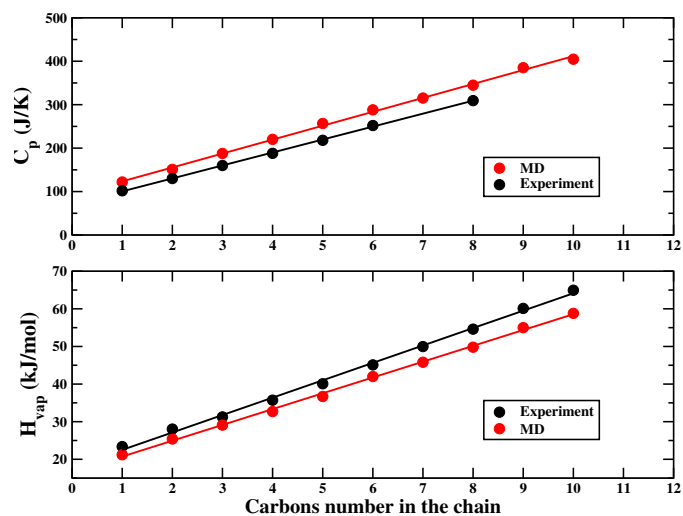}
		\caption{(Colour online) Top figure, heat capacity,  C$_p$, of the $n$-amines as function
			of the number of carbon atoms in the alkyl chains.
			Data calculated from this work (MD) and compared with
			experiments~\cite{35,36,37}. Bottom figure, enthalpy of vaporization
			as function of the number of carbon atoms in the alkyl chains.
			Data calculated from this work (MD) and compared with
			experiments~\cite{31}. Data at temperature $T = 298$~K.}
		\label{fig6}
	\end{center}
\end{figure}

Once the target properties were evaluated, the new parameters
were also tested
with other thermodynamic and dynamical quantities. In figure~\ref{fig6} and table~\ref{table4},
the heat capacities and enthalpies of vaporization were calculated
where we observed that the results were well compared to experiments.

\begin{table}[h]
	\begin{center}
		\caption{\small{Heat capacity (C$_p$), enthalpy of vaporization ($\Delta$H)
				and diffusion coefficients ($D$)
				for all the amines calculated from this work and the
				experiments.
				Data are for $T = 298$~K except for `*' which was calculated at $T = 259$~K.}}
		\label{table4}
		\begin{tabular}{|c|c|c|c|c|c|c|}
			\hline
			Amines & C$_{p}$ & C$_{p}$(exp) & $\Delta$H & $\Delta$H (exp) & D $\times 10^{-5}$
			& D  $\times 10^{-5}$ (exp) \\
			& (J/K) & (J/K) & (kJ/mol) & (kJ/mol) & (cm$^2$/s) & (cm$^2$/s) \\
			\hline
			Methylamine & 122.1$^*$ & 101.8$^*$ & 21.1 & 23.4 & 9.521 & 7.45 \\
			Ethylamine  & 151.1 & 130 & 25.4 & 28.0 & 8.654 & $-$    \\
			Propylamine & 187.7 & 160 & 29.1 & 31.3 & 3.039 & $-$    \\
			Butylamine  & 220.2 & 188 & 32.7 & 35.7 & 2.692 & $-$    \\
			Pentylamine & 256.8 & 218 & 36.7 & 40.1 & 2.392 & $-$    \\
			Hexylamine  & 288.1 & 252 & 42.0 & 45.1 & 1.785 & 1.55 \\
			Heptylamine & 315.2 &  $-$   & 45.8 & 49.9 & 1.672 & $-$    \\
			Octylamine  & 344.9 & 309 & 49.8 & 54.6 & 1.359 & 1.07 \\
			Nonylamine  & 385.2 &   $-$  & 55.0 & 60.1 & 1.123 & $-$    \\
			Decylamine  & 404.7 &  $-$   & 58.8 & 64.9 & 0.936 & 0.64 \\
			\hline
		\end{tabular}
	\end{center}
\end{table}

The heat capacity at constant pressure ($C_p$)
was calculated using the energy file obtained from the simulations in
the NPT ensemble described in section~\ref{sec2}. Then, the $C_p$ was evaluated from
GROMACS utilities with the defaults settings and
without the quantum corrections. For the  enthalpy of vaporization, the
following equation was used~\cite{34},
\begin{equation}
\Delta H_{\rm vap} = U_{\rm pot}({\rm gas}) - U_{\rm pot}({\rm liq}) + RT,
\label{3.3}
\end{equation}

\noindent $U_i$ is the potential energy, $R$ is the gas constant and $T$ is the
temperature. Here, the enthalpy was evaluated with simulations of
two boxes in the liquid and gas states. The liquid state was simulated
as described in section~\ref{sec2}, for the calculation of the dielectric constant.
For the gas state, it was assumed that the gas is ideal and
simulations were performed using a
stochastic dynamics integrator in the canonical ensemble as implemented
in the GROMACS software. The system contains one molecule and the simulations
were run for 10~ns after an equilibration period of 1~ns. Then, the
potential energy was obtained and then used to calculate the heat of vaporization.

\begin{figure}[h]
	\begin{center}
		\includegraphics[width=2.4in, angle=-90]{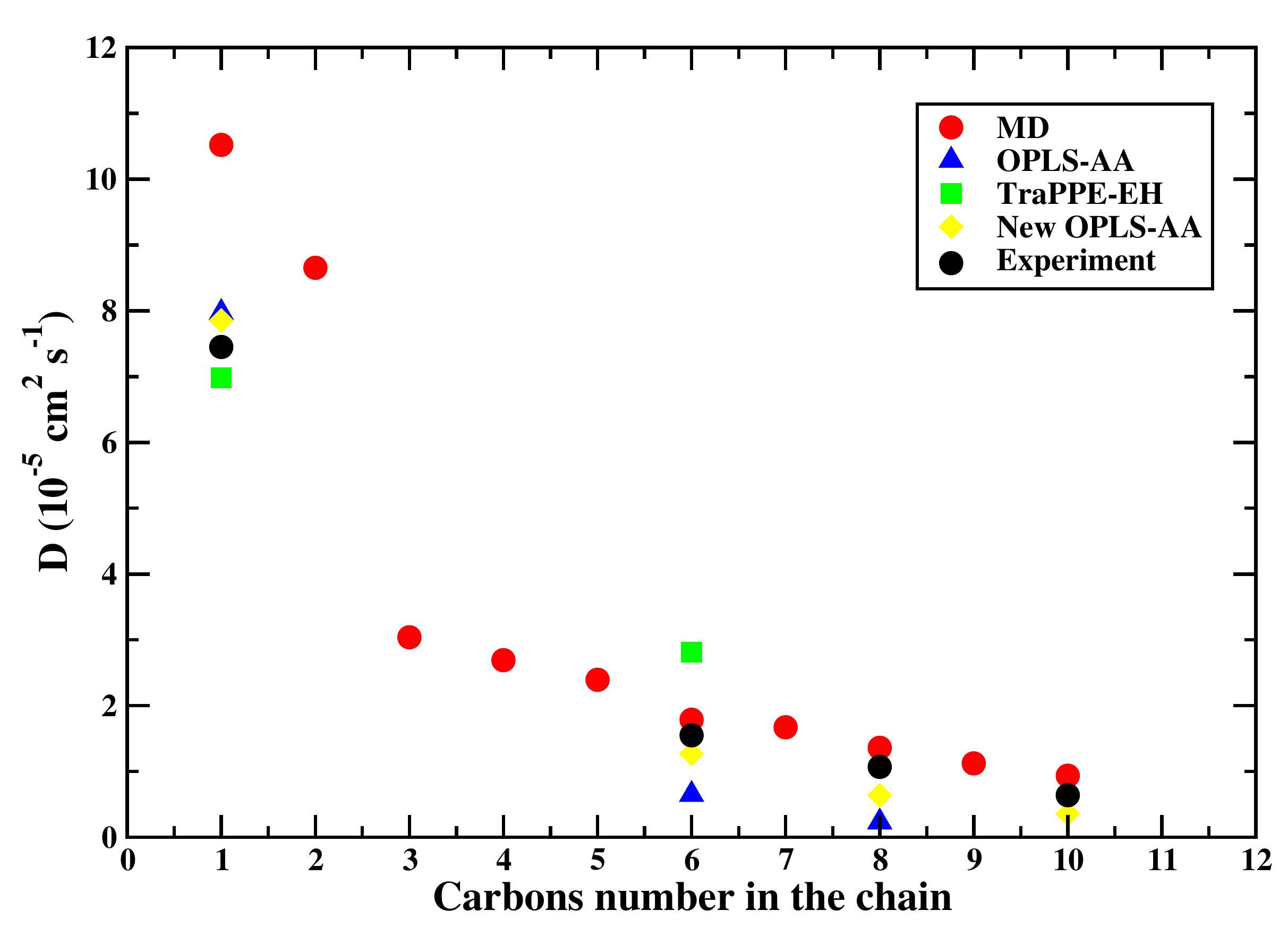}
		\caption{(Colour online) Diffusion coefficients of the $n$-amines as function
			of the number of carbon atoms in the alkyl chains at temperature $T = 298$~K.
			Data calculated from this work (MD) and compared with
			experiments.}
		\label{fig7}
	\end{center}
\end{figure}

The diffusion coefficients
were also calculated and plotted in figure~\ref{fig7} and in table~\ref{table4}.
They were evaluated with the Einstein relation, i.e.,  with the mean square
displacements, considering the linear region of the plots~\cite{28}.
\begin{equation}
D = \frac{1}{6t} \langle |r_i(t) - r_i(0)|^2 \rangle.
\label{3.4}
\end{equation}

\noindent The calculations of the diffusion coefficients were obtained over
the entire production time using the trajectory file and the defaults settings
of the GROMACS utilities for the separation of time origins. From table~\ref{table4} and
figure~\ref{fig7} we can see that the diffusion coefficients calculated in this
work have significant errors compared to the experiments~\cite{cas}.
Nevertheless, they have better agreement with them
than the other force fields reported in the literature~\cite{cas}.

\subsection{Miscibility}
From a previous work we know that miscibility of a solute
in water can be used as another target property to evaluate the force
field~\cite{38}. For the primary amines, we found that the short ones,
up to the pentylamine, are miscible in water whereas the long ones
are not~\cite{31}. Here,
in the present paper, miscibility was studied in terms of density
profiles, $\rho_s(z)$,
calculated from the NPT simulations in boxes
with homogeneous water/amine mixtures as initial configurations,
as described in section~\ref{sec2},
\begin{equation}
\rho(z) =  \frac{M*n(z)}{A \Delta Z},
\label{3.5} 
\end{equation}

\noindent where $n(z)$ is the number of amine molecules in a volume
of area $A$ and thickness $\Delta Z$. $M$ is the factor to convert the 
number density to mass density.

The density profiles were measured along the
$z$-axis using the GROMACS utility with 200~slices. Then, by analyzing
the profiles of both components, water and amines, it was
determined whether or not they were still mixing.

\begin{figure}[!h]
	\begin{center}
		\includegraphics[width=2.7in]{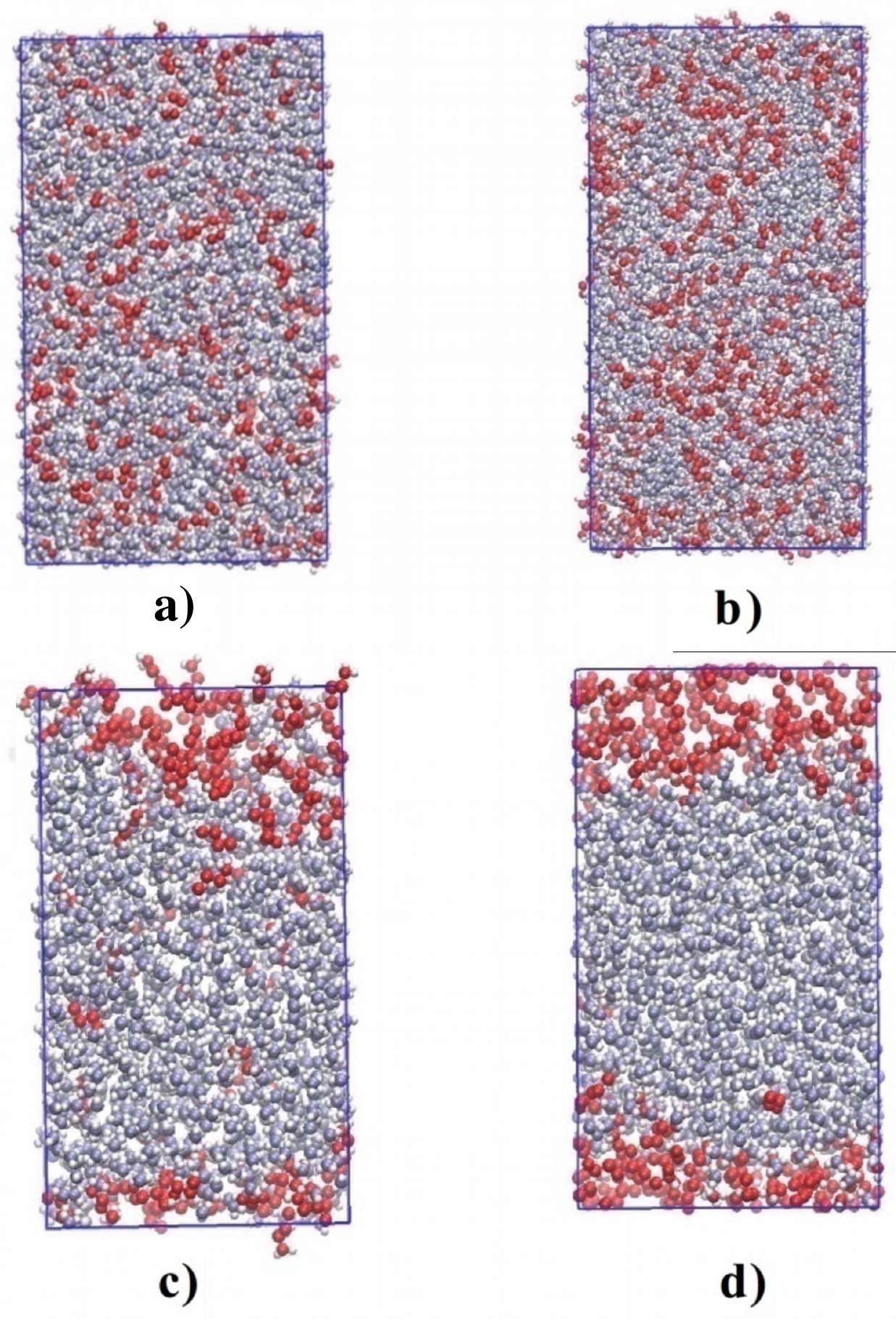}
		\caption{(Colour online) Snapshots of the last configuration of different
			$n$-amine/water mixtures. a)~Methylamine, b)~ethylamine,
			c)~propylamine and d)~butylamine. Water
			is represented in blue colour and amines in red colour.}
		\label{fig8}
	\end{center}
\end{figure}

\begin{figure}[!h]
	\begin{center}
		\includegraphics[width=2.8in, angle=-90]{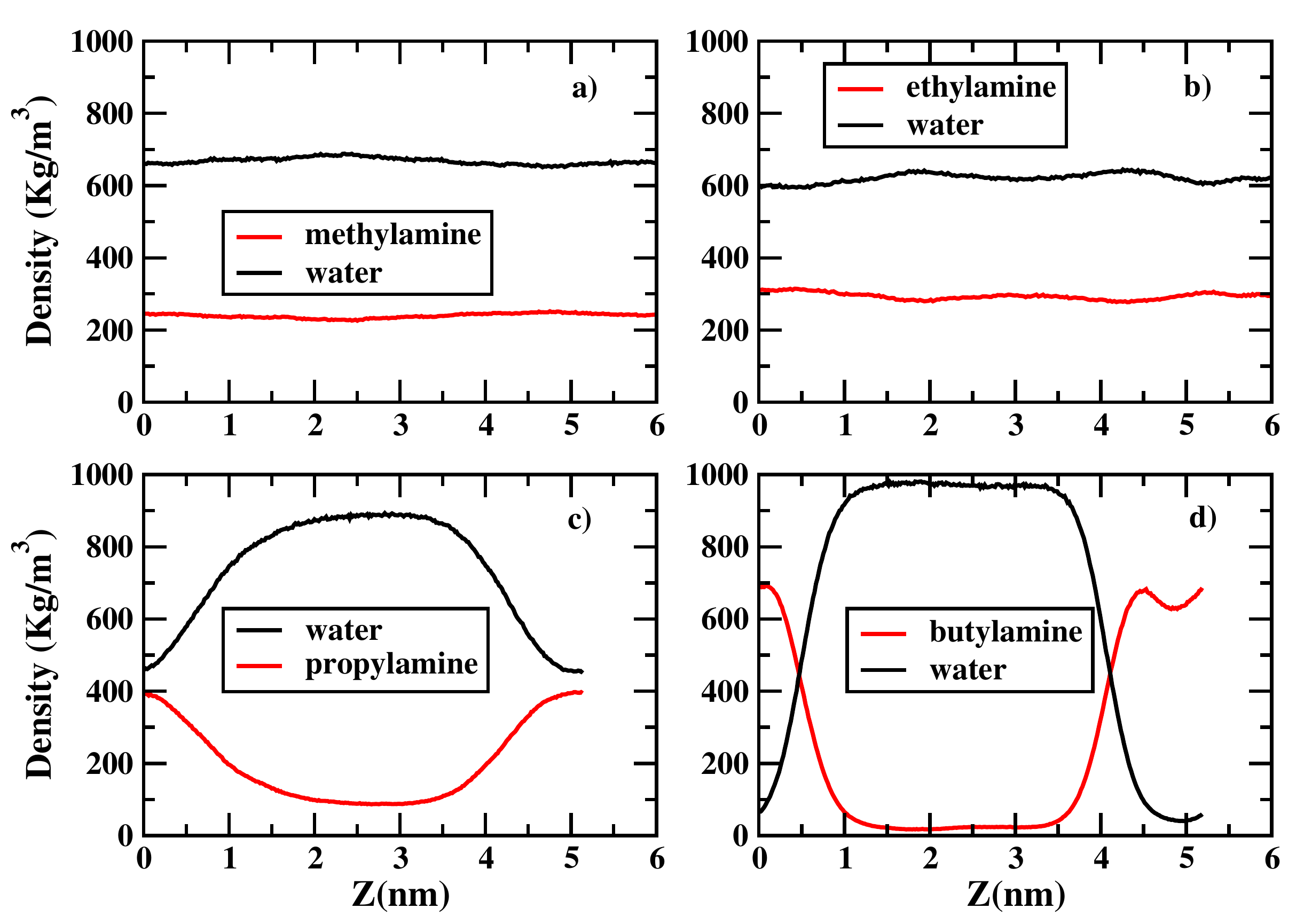}
		\caption{(Colour online) Density profiles, $\rho(z)$, for different water/$n$-amine mixtures,
			along the $z$-direction, at temperature $T = 298$~K.
			a)~Water/methylamine, b)~water/ethylamine, c)~water/propylamine and
			d)~water/butylamine. Black lines represent water and red lines represent
			the amine molecules.}
		\label{fig9}
	\end{center}
\end{figure}

The miscibility was evaluated from the amount of mass
of the solute, $m_s$, dissolved in water,
\begin{equation}
  m_s = m_w \frac{\rho_{s}(z)}{\rho_{w}(z)},
  \label{3.6} 
\end{equation}

\noindent where $m_w$ is the mass of the solvent, $\rho_s(z)$
and $\rho_w(z)$ are the density
profiles of the solute and water, respectively. Since the units
are given in g/l then, for a litter of water, $m_w$ is equal to 1000~g, i.e,
\begin{equation}
  Sol = 1000 \frac{\rho_{s}(z)}{\rho_{w}(z)} \,\, [\rm {g/l}].
   \label{3.7} 
\end{equation}
The results suggest that methylamine and ethylamine are
miscible in water
as can be seen in the snapshots of figures~\ref{fig8}a and~\ref{fig8}b, i.e., water and amines
are mixed in the whole simulation box. By calculating the
density profiles it is observed that water and the amines present uniform
distributions along the simulation box indicating that both systems are
well mixed (figures~\ref{fig9}a and~\ref{fig9}b)
in agreement with the experiments. From equation~\eqref{3.6}, the
values for the methylamine and ethylamine are 360~g/l and 471~g/l,
respectively, which suggest miscibility as reported in the literature for
those amines~\cite{31}. On the other hand, propylamine appears to be partial
miscible in water, only a few amine molecules are located in the bulk water
phase (figure~\ref{fig8}c) whereas butylamine looks immiscible
since two regions are well defined in the simulation box (figure~\ref{fig8}d).
The same conclusions can be stated from the density profiles.
For the water/propylamine system, two regions of high and small densities
are formed, suggesting that the two components are not completely mixed
(figure~\ref{fig9}c), i.e., there are regions rich in water or rich in amines.
In the case of the water/butylamine mixture, the system separates into
two well defined bulk phases as observed in the density profiles
of figure~\ref{fig9}d, there are observed regions
of nearly zero density for water and butylamine, i.e., the
system forms a liquid/liquid interface.
For long amines, the same trends are obtained
as butylamine (not shown here), i.e., those amines are not miscible in water.

\section{Conclusions}

In the present work molecular dynamics simulations were conducted to
construct a transferable force field for primary aliphatic amines. The
new force field was built up following the 3SSPP method where charges
and Lennard-Jones parameters were scaled to fit three target experimental
properties, dielectric constant, surface tension and density.

Different force fields have been reported in the literature that
calculate only a few thermodynamic or dynamical properties of amines.
In some cases they are not good enough when those quantities are
compared to actual experiments. Then,
in this work we intend to generalize a model that predicts
several experimental properties with good accuracy.
The procedure to obtain the new parameters of the
amine molecules is systematic and initially requires a reliable force field
for the subsequent reparameterization.

The results show that the new force field adequately reproduces
the target and other properties, although the shortest amines present
slightly larger errors
than the long ones compared to the experimental data. When miscibility
is evaluated, the results are good for the shortest and largest amines,
i.e., they show miscibility and no miscibility respectively, although for
the intermediate amines the results are not good enough, since they show
partial miscibility when they should be completely miscible
in water. Finally, since the present force field shows good agreement with
experiments it could be used as the starting one to characterize another type of
amines, such as secondary or tertiary amines, as well as branched and/or
aromatic amines, since they share an amino group with fixed partial charges.

\section{Acknowledgements}
This work was supported by DGAPA-UNAM-Mexico grant IN105120,
Conacyt-Mexico grant A1-S-29587 and DGTIC-UNAM LANCAD-UNAM-DGTIC-238
for supercomputer facilities.
HE-J and and ABS-A acknowledge the scholarship from Conacyt-Mexico.
We also acknowledge Alberto Lopez-Vivas, Cain Gonzales-Sanchez and Alejandro
Pompa for technical support.

\newpage

\ukrainianpart

\title{Розвиток нового методу силового поля для сімейства первинних аліфатичних амінів з використанням триетапної процедури систематичної параметризації}
\author{Х. Еспіноза-Хіменес, А. Б. Салазар-Арріага, Е. Домінгес}
\address{Інститут матеріалознавства, Національний автономний університет Мехіко, 04510, Мексика}

\makeukrtitle

\begin{abstract}
	\tolerance=3000%
		
У даній роботі оцінено застосовність триетапної процедури систематичної параметризації (3DSSPP) для розвитку методу силового поля для первинних амінів. Попереднє моделювання первинних амінів показує, що в кімнатних умовах силові поля потоків можуть занижувати деякі експериментальні значення. Тому ми пропонуємо новий набір параметрів для моделі об’єднаного атома, який можна використовувати для коротких і довгих амінів, що дозволяє точно спрогнозувати термодинамічні та динамічні властивості. З дотриманням методології 3SSPP, часткові заряди вибираються так, щоб відтворити експериментальні значення діелектричної константи, тоді як параметри Леннарда-Джонса, $\epsilon$ та $\sigma$, підбираються для відтворення поверхневого натягу на межі ``пара-рідина'' та густини рідини, відповідно.
Спочатку було проведено моделювання для молекули пропіламіну шляхом введення трьох різних типів атомів вуглецю: C$_\alpha$ та C$_\beta$ з електричними зарядами і C$_n$ без заряду. Потім, міняючи заряди вуглеців і використовуючи змінні параметри Леннарда-Джонса, було знайдено новий набір констант для довгих амінів.	
Результати показують хорошу узгодженість з експериментальними значеннями діелектричної константи та густини з похибкою менше 1\%, тоді як для поверхневого натягу похибка зростає до 4\%. Для коротких амінів, метиламіну та етиламіну, нові заряди були отримані з підгонки, зробленої на основі результатів для довгих амінів. Для цих молекул значення діелектричної константи та поверхневого натягу мають похибки порядку 10\% у порівняннні з експериментальними даними. Змішуваність амінів також була перевірена з використанням нових параметрів, і результати показали хорошу узгодженість з експериментом.
	\keywords аміни, силове поле, параметри Леннарда-Джонса, заряди, молекулярна динаміка
	
\end{abstract}

\lastpage
\end{document}